\documentclass{natureprintstyle}
\usepackage{epsfig,caption}
\usepackage{color}
\usepackage{bm}
\usepackage{graphicx}
\usepackage{longtable}
\usepackage{amssymb}
\usepackage{rotating}
\usepackage{latexsym}
\usepackage{hyperref}

\newcommand{\apj}{Astrophys. J.}

\newcommand{\pasp}{Publ. Astron. Soc. Pac.}
\newcommand{\apjs}{Astrophys. J. Supp.}

\newcommand{\mnras}{Mon. Not. R. Astron. Soc.}
\newcommand{\apjl}{Astrophys. J. Let.}
\newcommand{\aap}{Astron. Astrophys.}
\newcommand{\aj}{Astron. J.}
\newcommand{\nat}{Nature}

\title{A massive, quiescent, population II galaxy at a redshift of 2.1}

\author{Mariska Kriek$^{1}$, Charlie Conroy$^2$, Pieter G. van Dokkum$^3$, Alice E. Shapley$^{4}$, Jieun Choi$^2$, Naveen A. Reddy$^{5}$,  Brian Siana$^{5}$, Freeke van de Voort$^{1}$, Alison L. Coil$^{6}$, Bahram Mobasher$^{5}$}

\begin{document}

\maketitle

\let\thefootnote\relax\footnote{
\begin{affiliations}
\item Department of Astronomy, University of California, Berkeley, CA 94720, USA 
\item Department of Astronomy, Harvard University, Cambridge, MA, USA
\item Astronomy Department, Yale University,  New Haven, CT, USA
\item Department of Physics \& Astronomy, University of California, Los Angeles, CA 90095, USA
\item Department of Physics \& Astronomy, University of California, Riverside, CA 92521, USA
\item Center for Astrophysics and Space Sciences, University of California, San Diego, La Jolla, CA 92093, USA

\end{affiliations}
}

\begin{abstract}
Unlike spiral galaxies such as the Milky Way, the majority of the
stars in massive elliptical galaxies were formed in a short period
early in the history of the Universe. The duration of this formation
period can be measured using the ratio of magnesium to iron
abundance ([Mg/Fe])\cite{FMatteucci1994,STrager2000,DThomas2005,CConroy2014},
which reflects the relative enrichment by core-collapse and type Ia
supernovae. For local galaxies, [Mg/Fe] probes the combined formation
history of all stars currently in the galaxy, including younger and
metal-poor stars that were added during late-time
mergers\cite{PvanDokkum2010}. Therefore, to directly constrain the
initial star-formation period, we must study galaxies at earlier
epochs. The most distant galaxy for which [Mg/Fe] had previously been
measured\cite{ILonoce2015} is at a redshift of $z\approx1.4$, with
[Mg/Fe]~$=0.45^{+0.05}_{-0.19}$. A slightly earlier epoch ($z\approx
1.6$) was probed by stacking the spectra of 24 massive quiescent
galaxies, yielding an average [Mg/Fe] of
$0.31\pm0.12$\cite{MOnodera2015}. However, the relatively low
signal-to-noise ratio of the data and the use of index analysis
techniques for both studies resulted in measurement errors that are
too large to allow us to form strong conclusions. Deeper spectra at
even earlier epochs in combination with analysis techniques based on
full spectral fitting are required to precisely measure the abundance
pattern shortly after the major star-forming phase ($z>2$). Here we
report a measurement of [Mg/Fe] for a massive quiescent galaxy at a
redshift of $z=2.1$, when the Universe was 3 billion years old. With
[Mg/Fe]~$=0.59\pm0.11$, this galaxy is the most Mg-enhanced massive
galaxy found so far, having twice the Mg enhancement of similar-mass
galaxies today. The abundance pattern of the galaxy is consistent with
enrichment exclusively by core-collapse supernovae and with a
star-formation timescale of 0.1 to 0.5 billion years --
characteristics that are similar to population II stars in the Milky
Way. With an average past star-formation rate of 600 to 3,000
solar masses per year, this galaxy was among the most vigorous
star-forming galaxies in the Universe.
\end{abstract}

\begin{figure}[!t]
\begin{center}
\includegraphics[width=0.48\textwidth]{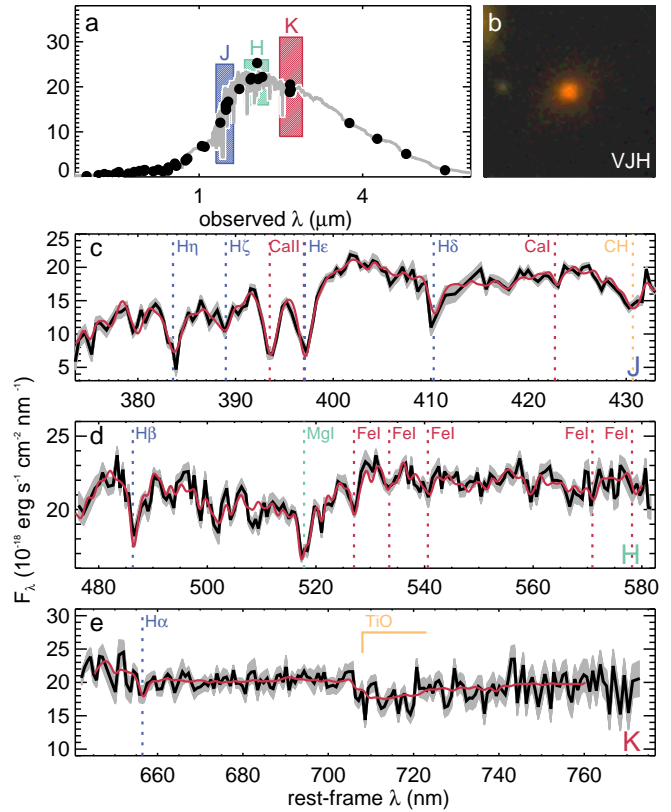}
\caption{{\bf Photometry, image and MOSFIRE spectrum of
    COSMOS-11494.} {\bf a,} Multi-wavelength spectral energy
  distribution (black circles) and best-fitting stellar population model to
  only the photometry (grey line). {\bf b,} {\it HST} colour (V, J, and H)
  image. {\bf c-e,} MOSFIRE spectrum in three wavelength intervals (J,
  H, and K; black), corresponding to the coloured areas in {\bf
    a}. The gray shaded regions represent the 1\,$\sigma$ uncertainty
  on the flux. The best-fitting stellar population model used to derive
  the age and abundance pattern is shown in red.\label{fig:spec}}
\end{center}
\end{figure} 

We observed the galaxy COSMOS-11494 with the near-infrared
multi-object spectrograph MOSFIRE on the \textit{Keck~I
  Telescope}\cite{IMcLean2012}. It was also observed by two other
programmes\cite{SBelli2014,MKriek2015}, and so we incorporated these
publicly available archival data. COSMOS-11494 was selected from the
3D-HST survey\cite{RSkelton2014,IMomcheva2015}. With a stellar mass
$M$ given by ${\rm log_{10}}\,M/M_\odot=11.5\pm0.1$, COSMOS-11494 is
among the most massive galaxies at its redshift, and it has a very low
star-formation rate of less than $0.6M_\odot/$~yr (see
Methods). Similarly to the typical massive, quiescent galaxy at this
redshift, it is smaller than its local counterparts of the same mass,
with an effective radius of 2.1~kpc\cite{JvandeSande2013}. The MOSFIRE
rest-frame optical spectrum, the multi-wavelength spectral energy
distribution\cite{RSkelton2014} and \textit{Hubble Space Telescope}
(\textit{HST}) colour image\cite{RSkelton2014} of COSMOS-11494 are
shown in Figure~\ref{fig:spec}.

\begin{figure*}[!t]
\begin{center}
\includegraphics[width=0.98\textwidth]{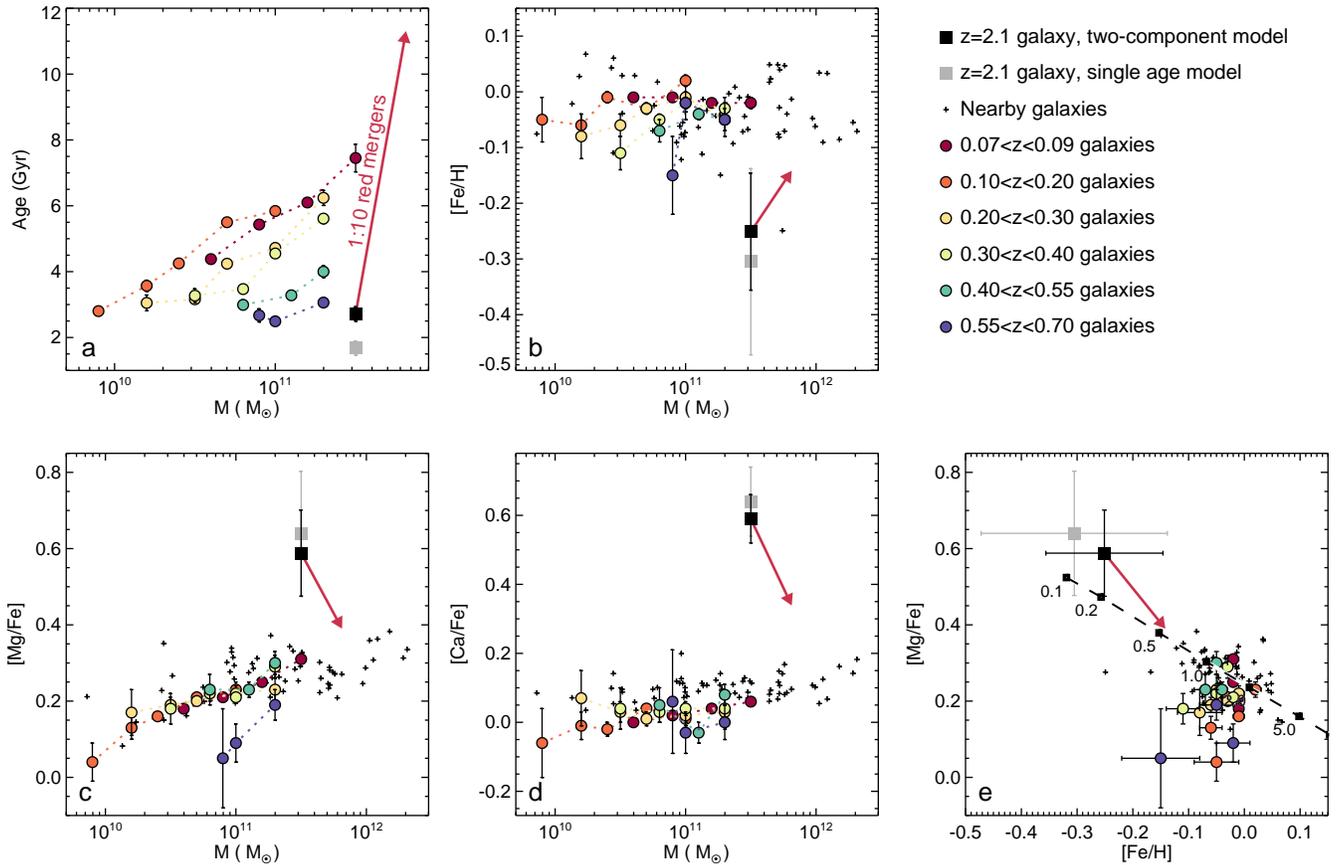}
\caption{\textbf{Age and abundance patterns of COSMOS-11494 in
    comparison to lower-redshift quiescent galaxies.} \textbf{a,}
  Stellar population age, \textbf{b,} [Fe/H], \textbf{c,} [Mg/Fe], and
  \textbf{d,} [Ca/Fe] versus stellar mass. \textbf{e,} [Mg/Fe] versus
         [Fe/H]. The black dashed line represents  a chemical
         evolution model for different star-formation timescales in
         Gyr. In all panels, the black and grey filled squares
         represent COSMOS-11494 for the two-component and single-age
         model, respectively, the coloured symbols represent
         low-redshift galaxies\cite{JChoi2014} binned by mass and
         redshift, and the small black pluses are nearby
         galaxies\cite{CConroy2012d}. Error bars are 1\,$\sigma$. The
         red arrows represent the simple evolutionary model (see main
         text). \label{fig:abun}}
\end{center}
\end{figure*}

Here we measure the stellar abundance pattern of COSMOS-11494 from the
MOSFIRE rest-frame optical spectrum with our absorption line fitter
(\texttt{alf}) code\cite{CConroy2012a} (see Methods). For our default
model we adopt a two-component stellar population, for which the age
of both components and the slope of the stellar initial mass function
(IMF) are free parameters. To enable comparison with previous
work\cite{JChoi2014}, we also fit the spectrum with a single-age model
and a Kroupa\cite{PKroupa2001} IMF.

For the default model we find ${\rm [Fe/H]}=-0.25\pm 0.11$, ${\rm
  [Mg/Fe]}=0.59\pm0.11$, ${\rm [Ca/Fe]}=0.59\pm0.07$ and an age of
$2.71\pm0.22$ Gyr. The best-fitting mass-to-light ratio ($M/L$) is
consistent with the $M/L$ assuming a Kroupa IMF ($(M/L)/(M/L_{\rm
  Kroupa})=0.97\pm0.55$), although the error is large because of the
insufficient S/N of the spectrum and the lack of rest-frame
near-infrared coverage. We also fit this model with $\lambda_{\rm
  restframe}<4000$~\AA\ excluded, and find similar values. For the
single-age model we find similar abundance ratios as for the
two-component model, but the modelled age is 1~Gyr younger. This
difference is expected, because younger stellar populations have lower
$M/L$ and so have larger weights in the fit. 

In Figure~\ref{fig:abun} we compare the spectral modelling results of
COSMOS-11494 with those of galaxies at $0.05<z<0.7$\cite{JChoi2014}
and of a sample of nearby massive galaxies\cite{CConroy2012d}. All
galaxies are fitted with the \texttt{alf} code. Figure~\ref{fig:abun}
illustrates that COSMOS-11494 is more Mg-enhanced than similar-mass
galaxies at lower redshift, with [Mg/Fe] about 0.3~dex higher. [Ca/Fe]
is also higher compared to the values for lower-redshift massive
galaxies.

To interpret the abundance pattern of COSMOS-11494, we show a chemical
evolution model in Figure~2e, which assumes a Salpeter
IMF\cite{ESalpeter1955}, a constant star-formation history over a
given timescale, a core-collapse\cite{CKobayashi2006} and a type Ia
supernova yield model\cite{KNomoto1984}; we also adopt a  power-law
delay-time distribution of the from $t^{-1}$ for type Ia supernovae
that occured between 0.1 and 13~Gyr\cite{DMaoz2012}. The
star-formation timescale decreases along the curve, with the highest
value of [Mg/Fe] corresponding to the shortest timescale of
0.1~Gyr. The relatively low Fe abundance in combination with the high
[Mg/Fe] favours a short star-formation timescale of around
$0.2$~Gyr. Therefore, this model implies that COSMOS-11494 has
experienced very little enrichment by type Ia supernovae.

However, the best-fitting timescale strongly depends on the assumed
delay time of prompt type Ia supernovae. This parameter is poorly
constrained in models and depends on the type Ia progenitor
model\cite{CKobayashi2009}; for the double degenerate scenario the
lifetime can be as short as about 0.1~Gyr, whereas for a single
degenerate scenario it can be as high as about
0.5~Gyr\cite{CKobayashi2009}. However, the delay times of  prompt type
Ia supernovae inferred from observations are as short as
$0.1$~Gyr\cite{DMaoz2012}, and so 0.5~Gyr may be a conservative upper
limit. The assumed IMF affects the chemical evolution model as well,
and a flatter IMF results in a longer timescale. Finally, the adopted
chemical evolution model depends on the core-collapse supernova
yields\cite{CKobayashi2006}, and other yield models were unable to
reproduce the observed high [Mg/Fe] in combination with the observed
[Fe/H]\cite{FThielemann1996,SWoosley1995}. Nonetheless, individual
stars with similarly high [Mg/Fe] and [Fe/H] values as COSMOS-11494
have been identified in the bulge of the Milky
Way\cite{JFulbright2007}, which supports the validity of the adopted
yield model\cite{CKobayashi2006}. Taking into account all
uncertainties on our chemical evolution model, we estimate a
star-formation timescale of $\sim0.1-0.5\,$Gyr. 

Ca, which is also produced and returned to the interstellar medium
through core-collapse supernovae, also shows a strong enhancement with
respect to Fe; the difference compared to low-redshift analogues is
even more extreme than for Mg.  This differential evolution of Ca and
Mg with time is unexpected, because both elements are formed in
massive stars, but metallicity-dependent core-collapse-supernova
yields might explain the differences\cite{CKobayashi2006}. 

When combining the star-formation timescale and the best-fitting
stellar mass, we find an average past star-formation rate of
(600--3,000)\,$M_\odot$/yr. The single-age best-fitting SPS model sets
a lower limit on the formation redshift of $z>4$. For the
two-component model we find an average formation redshift of
$z=12^{+9}_{-4}$. The inferred star-formation rate and formation epoch
are consistent with the properties that were derived for the most
active galaxy found so far, HFLS3\cite{DRiechers2013}. This dusty
sub-millimetre galaxy has a star-formation rate of 2,900 $M_\odot$/yr
and a redshift of $z=6.34$, and so could be similar to the
star-forming progenitor of COSMOS-11494.

\begin{figure}
\begin{center}
\includegraphics[width=0.35\textwidth]{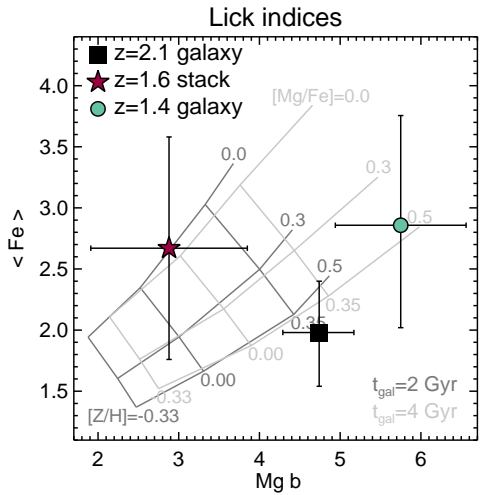}
\caption{{\bf Abundance pattern of COSMOS-11494 in comparison to other
    high-redshifts galaxies.} We show the Lick indices
  $\langle$Fe$\rangle$ and Mg~b for COSMOS-11494, a $z\approx1.4$
  quiescent galaxy\cite{ILonoce2015} observed with {\it
    VLT}/X-Shooter, and a stack of $z\approx1.6$ quiescent galaxy
  spectra\cite{MOnodera2015} observed with {\it Subaru}/MOIRCS. Error
  bars are 1\,$\sigma$. We also show grids based on SPS
  models\cite{DThomas2003} for a range of range of values of [Z/H] and
  [Mg/Fe]. The dark and light grey grids are for galaxy ages of 2~Gyr
  and 4~Gyr, respectively.\label{fig:lick}}
\end{center}
\end{figure}

COSMOS-11494 seems to be more Mg-enhanced than $z\approx1.5$
galaxies in previous work. However, because different methods have been
used in these studies, systematic differences may occur, and the
derived values cannot be directly compared\cite{CConroy2014}. For a
more direct comparison to previous studies, we use the Lick
indices\cite{GWorthey1994} $\langle$Fe$\rangle$ (=[Fe5270+Fe5335]/2)
and Mg\,$b$. For the $z=1.4$ galaxy, the spectrum was not deep enough
to measure the two Fe lines needed to determine
$\langle$Fe$\rangle$. Instead, we calibrated $\langle$Fe$\rangle$ for
two other iron lines (Fe4388 and Fe5015) using a set of SPS
models\cite{DThomas2003}, and derived $\langle$Fe$\rangle$ from these
(marginally detected) lines for the $z=1.4$
galaxy. Figure~\ref{fig:lick} shows the measurements in comparison to
a grid of SPS models\cite{DThomas2003} for a range of metallicities
([Z/H]$=-0.33, 0.0, 0.35$) and [Mg/Fe] values (0.0, 0.3, 0.5).

For COSMOS-11494 the [Mg/Fe]~$\approx 0.6$ implied from the Lick
indices is consistent with the modelling results. This value
illustrates that COSMOS-11494 -- which probes an earlier epoch than
previous work -- is indeed more Mg-enhanced than the quiescent galaxies
in the other studies. Furthermore, our high S/N spectrum results in
the most robust abundance pattern measurement for a distant galaxy so
far. Although the two different approaches to derive the abundance
pattern give consistent results for COSMOS-11494, for distant
galaxies, the full spectral modelling approach is strongly preferred
over the approach involving Lick indices (Methods).

The high [Mg/Fe] of COSMOS-11494 compared to lower-redshift quiescent
galaxies of similar mass suggests that this galaxy, and possibly other
distant quiescent galaxies, do not passively evolve into quiescent
early-type galaxies today. A similar conclusion was drawn from the
small sizes of distant quiescent galaxies compared to their local
analogs\cite{PvanDokkum2008}, and in the past several years it has
become apparent that distant quiescent galaxies grow in mass and size
by accreting primarily smaller
galaxies\cite{TNaab2009,PvanDokkum2010}. This inside-out growth by
late-time mergers with less-massive galaxies predicts a decline in
[Mg/Fe], because lower-mass galaxies are less
Mg-enhanced\cite{DThomas2005,JChoi2014}

In Figure~\ref{fig:abun} we explore this scenario by showing the
predicted path of COSMOS-11494 for a simple evolutionary model. We
assume that the galaxy grows by red minor (1:10) mergers with smaller,
less Mg-enhanced quiescent galaxies with [Mg/Fe]~$=0.18$,
[Fe/H]~$=-0.05$, and [Ca/Fe]~$=0.02$\cite{JChoi2014}, and that the
mass nearly doubles between $z=2.1$ and the present
day\cite{PvanDokkum2010} following the mass evolution ${\rm
  d(log}\,M)/{\rm d}z=-0.15$. Therefore, we assume no evolution in
[Mg/Fe] and [Ca/Fe] at lower masses. If the abundance ratios for these
galaxies would be higher at earlier times as well, then the predicted
evolution would be less strong. To estimate the evolution in galaxy
age for the merger model, we assume that the age is proportional to
$M^{0.3}$, as was found for $z<0.7$ galaxies\cite{JChoi2014}. 

Figure~\ref{fig:abun} shows that the merger model can substantially
decrease [Mg/Fe] and [Ca/Fe], and increase [Fe/H]. However, there are
several caveats to our simple model comparison. First, the
$0.07<z<0.7$ measurements are derived by fitting a single-age model,
and so are sensitive to low levels of recent star formation. Second,
we assume that the accreted stars are well-mixed with the {\it
  in-situ} population. However, simulations of galaxy formation show
that the added material is mostly deposited in the outskirts of the
galaxy\cite{TNaab2009}, and so the net evolution due to mergers -- in
the central parts targeted by the spectrographs -- may be less. Third,
it is unlikely that the descendent of COSMOS-11494 is a typical
massive, quiescent galaxy today. The star formation in many
low-redshift quiescent galaxies is quenched at later times, resulting
in longer star-forming periods and, hence, lower [Mg/Fe]. Therefore,
the descendant of COSMOS-11494 presumably resides in the tail of the
low-redshift distributions. Finally, the model does not include
possible late-time star formation or mergers with star-forming
galaxies, which would also result in a decrease in [Mg/Fe] with time. 

More spectra of quiescent galaxies at high redshifts are needed to
measure the evolution of the slope and the intercept of the age\,$-M$
and [Mg/Fe]\,$-M$ relation. These measurements could eventually
discriminate between different evolutionary scenarios, and the amount
of mixing of stars after galaxy mergers\cite{JChoi2014}. In
combination with more accurate supernova progenitor and yield models,
and therefore improved chemical evolution models, these measurements
will also provide unique information on the star-formation histories
of the most massive galaxies and their possible role in the
reionization of the Universe at $z>6$. We expect that observations
with NIRspec on the \textit{James Webb Space Telescope} will
revolutionize this field within the next five years, with future
ultra-deep observations with MOSFIRE paving the way.


\begin{addendum}
 \item[Acknowledgements] M. K. acknowledges discussions with
   J. Greene and E. Quataert. The data presented in this paper were
   obtained at the W.\,M. Keck Observatory, which is operated as a
   scientific partnership among the California Institute of
   Technology, the University of California and the National
   Aeronautics and Space Administration. The Observatory was made
   possible by the generous financial support of the W.M. Keck
   Foundation. The authors wish to recognize and acknowledge the very
   significant cultural role and reverence that the summit of Mauna
   Kea has always had within the indigenous Hawaiian community. We are
   most fortunate to have the opportunity to conduct observations from
   this mountain. We acknowledge support from NSF AAG collaborative
   grants AST-1312780, 1312547, 1312764, and 1313171 and archival
   grant AR-13907, provided by NASA through a grant from the Space
   Telescope Science Institute. C.C. acknowledges support from NASA
   grant NNX13AI46G, NSF grant AST-1313280, and the Packard
   Foundation.
 \item[Correspondence] Correspondence and requests for materials
should be addressed to M. K. ~(email: mkriek@berkeley.edu).
\item[Author Contributions] M. K., P. G. v. D. \& C. C., wrote the primary Keck proposal. M. K. \& C. C. led the interpretation. M. K. wrote the reduction pipeline, reduced the data, determined the stellar mass, measured the Lick indices, and wrote the text. C. C. developed the SPS model, fitted the spectrum, and derived the chemical evolution model. M. K., P. G. v. D., J. C., F. v. d. V., and N. A. R. did the observations. All authors contributed to the analysis and interpretation.
\end{addendum}

\clearpage
\newpage

\begin{center}
{\bf \Large \uppercase{Methods} }
\end{center}

\textbf{Best-fitting model to the photometry.} 
To derive the stellar mass of COSMOS-11494, we fit the
broadband photometry with the flexible stellar population synthesis (SPS)
models\cite{CConroy2009,CConroy2010}. We assume a delayed exponential
star formation history of the form SFR~$\propto t\,e^{-t/\tau}$ and
the dust attenuation law from Kriek \& Conroy\cite{MKriek2013}. We adopt
the Chabrier stellar IMF\cite{GChabrier2003} to facilitate direct
comparison with lower-redshift studies\cite{JChoi2014}. 

The error bar on the stellar mass is completely dominated by
systematic uncertainties. We estimate this uncertainty by varying the
SPS model\cite{GBruzual2003,CMaraston2005}, dust attenuation law
\cite{JCardelli1989,DCalzetti2000}, parameterization of the
star-formation history, and the scaling of the broadband spectral
energy distribution. We do not vary the IMF, as it can be approximated
by a simple offset in the stellar mass. A Kroupa IMF\cite{PKroupa2001}
would have resulted in a similar stellar mass as for the Chabrier IMF,
whereas a Salpeter IMF\cite{ESalpeter1955} would have resulted in a
stellar mass a factor of 1.6 higher.

\textbf{Spectral fitting.} Parameters are estimated from the
rest-frame optical spectrum with the
\texttt{alf}\cite{CConroy2012a,CConroy2014,JChoi2014} code. This code
combines libraries of isochrones and empirical stellar spectra with
synthetic stellar spectra covering a wide range of elemental abundance
patterns. The code fits for C, N, O, Na, Mg, Ca, Ti, V, Cr, Mn, Fe,
Co, Ni, redshift, velocity dispersion, and several emission lines. The
stellar population age and IMF are free parameters as well, and
multiple stellar population components are allowed. When fitting a
two-component model, the age represents the mass-weighted average age
of the two separate components. The ratio of the model and data are fitted
by a high order polynomial to avoid potential issues with the
flux calibration of the data. The fitting is done using a Markov chain
Monte Carlo algorithm\cite{DForeman-Mackey2013}.

\textbf{Lick indices versus full spectral modelling.}  As mentioned in
the main text, for COSMOS-11494 the [Mg/Fe]~$\approx0.6$ implied from
the Lick indices is consistent with the modelling results. The
metallicity measurements also agree between the two methods: our
best-fitting values for [Mg/Fe] and [Fe/H] imply [Z/H]$=0.25$
(=[Fe/H]+0.94[Mg/Fe]\cite{DThomas2003}), which is consistent with the
model shown in Figure~\ref{fig:lick} for the best-fitting age of
2.5\,Gyr. For the $z\approx1.4$ individual galaxy and the $z\sim1.6$
stack, the abundance patterns are based on the Lick indices, and so
by construction they should be closer to the grid points. This is
indeed the case for $z\approx1.4$ individual galaxy\cite{ILonoce2015}
with a [Mg/Fe] of $0.45^{+0.05}_{-0.19}$ and a [Z/H] of
$0.61^{+0.06}_{-0.05}$. For the derived age of 4\,Gyr the two Lick
indices are consistent with the model grid. However, 
the individual Fe lines yield very different and inconsistent
results when calculating $\langle \rm Fe \rangle$. For the $z\approx1.6$
galaxy stack\cite{MOnodera2015}, the derived
[Mg/Fe]~$=0.31^{+0.12}_{-0.12}$, [Z/H]~$={0.24^{+0.20}_{-0.14}}$ and
log$(\rm age/Gyr)=0.04^{+0.10}_{-0.08}$ are less consistent with the
derived values based on all Lick indices, though the error bars are
large. 

Although the two different approaches to derive the abundance pattern
agree well for COSMOS-11494, the full spectral modelling approach is
strongly preferred over the appraoch involving Lick indices. The
rest-frame optical spectrum of $z>1$ galaxies has been shifted to
near-infrared wavelengths. The many skylines at these wavelengths in
combination with the relatively lower S/N of the
spectra of distant galaxies result in large error bars on the
measurement of a single feature. Lick indices are integrated
measurements and do not take into account wavelength-dependent
features within the bandpass. Therefore, skylines will severely complicate
their measurement because affected wavelengths are not
down-weighted. Furthermore, skylines result in non-Gaussian and
correlated noise properties, and so error bars on Lick indices are
usually underestimated.  Consequently, Lick indices are much more prone
to systematic errors.  The discrepancies between individual Lick
indices and the derived stellar abundance pattern based on all Lick
indices for the two lower-redshift measurements further illustrate
this point. By modelling the full spectrum, we use many more features
and can better deal with the larger uncertainties in regions affected
by skylines.

\textbf{Code availability.} The data reduction package used to
process the raw MOSFIRE data will be made public in the coming year at
http://astro.berkeley.edu/\~\,mariska. To derive the stellar
mass, we used the flexible SPS models which are available at
https://github.com/cconroy20/fsps and the SPS fitting code
\texttt{FAST}\cite{MKriek2009b}, which is publicly available at
http://astro.berkeley.edu/\~\,mariska/fast/. The spectral fitting code
\texttt{alf}\cite{CConroy2012a,CConroy2014,JChoi2014} that was used to derive
the abundance pattern is not publicly available, but the underlying
model components are available for download from
http://scholar.harvard.edu/cconroy/sps-models.

\textbf{Data availability.} The one-dimensional original and binned
spectrum shown in Figure~\ref{fig:spec} and corresponding 1\,$\sigma$
uncertainties, as well as the best-fitting model spectrum are
available as Source Data. The binned spectrum is constructed by first
masking wavelengths affected by skylines and poor atmospheric
transmission, and then taking the median of the flux of ten non-masked
consecutive pixels. The photometric data points shown in
Figure~\ref{fig:spec} are made available by the 3D-HST collaboration
(catalog version v4.1) at http://3dhst.research.yale.edu/Data.php. The
abundance pattern for COSMOS-11494 for both the two-component and
single-age model, as shown in Figure~\ref{fig:abun}, is also available
as Source Data.

\end{document}